# The Oasis Project: UHI mitigation strategies applied to Parisian schoolyards

*Ghid Karam[1]*
*Univ Paris Diderot, Sorbonne Paris Cité, LIED, UMR 8236, CNRS, F-75013, Paris, (France)*
*Université Paris-Est, ESIEE Paris, département SEN, F-93162, Noisy-le-Grand, France*

*Martin Hendel*
*Univ Paris Diderot, Sorbonne Paris Cité, LIED, UMR 8236, CNRS, F-75013, Paris, (France)*
*Université Paris-Est, ESIEE Paris, département SEN, F-93162, Noisy-le-Grand, France*

*Cécilia Bobée*
*Univ Paris Diderot, Sorbonne Paris Cité, LIED, UMR 8236, CNRS, F-75013, Paris, (France)*

*Alexandre Berthe*
*Univ Paris Diderot, Sorbonne Paris Cité, LIED, UMR 8236, CNRS, F-75013, Paris, (France)*

*Patricia Bordin*
*Univ Paris Diderot, Sorbonne Paris Cité, LIED, UMR 8236, CNRS, F-75013, Paris, (France)*

*Laurent Royon*
*Univ Paris Diderot, Sorbonne Paris Cité, LIED, UMR 8236, CNRS, F-75013, Paris, (France)*

## ABSTRACT

Paris is experimenting and implementing strategies to increase the capital's resilience and promote climate change adaptation. Major heatwaves have been hitting the French Capital lately (Bador et al., 2017) and have thus focused attention on UHI countermeasures, such as pavement-watering solutions and urban greening. The Oasis Project is one such strategy aiming to transform schoolyards into urban cool islands that would benefit surrounding neighborhoods and their inhabitants during heatwaves.

The work presented here focuses on identifying high-priority schoolyards among the 670 city-owned schools. This is conducted using a GIS tool used to identify areas with high cooling potential, which would benefit most from UHI countermeasures. After extracting bare schoolyard from the facilities, we built a cooling indicator based on the ratio of high solar irradiance surface to the whole schoolyard area. We were thus able to identify 38 schoolyards with high cooling potential, 157 with medium cooling potential and 286 facilities with moderate cooling potential, out of the 670 facilities.

The methodology can be applied to other cities, and therefore helps set up GIS tools that can provide municipalities with meaningful insight into their urban cooling strategy.

---

[1] Ghid Karam: ghid.karam@esiee.fr.

# Introduction

Climate change is expected to result in an overall increase in annual temperatures in France: 1 to 4°C compared to the current reference (12.4°C) (Bador et al., 2017) as well as a multiplication of heat waves, with frequencies of up to 25 heat-wave days per year compared to 1 day at present (Lemonsu, Kounkou-Arnaud, Desplat, Salagnac, & Masson, 2013). These heat waves are amplified by the urban heat island (UHI) phenomena rooted in several different city parameters: urban morphology, permeability, as well as human activity (Oke, 1982)

The 2003 heat wave caused 70,000 deaths in Western Europe, and particularly affected Paris where its impacts were aggravated by the UHI effect (Robine et al., 2008).

The school network managed by the Paris City Hall is a dense network of 656 kindergarten and elementary schools and 115 middle schools. These areas cover nearly 73 hectares, which tallies to nearly 14% of the actual areas of parks and gardens within the City[2], and are evenly distributed throughout the capital, located within a 200 m radius of each inhabitant. For the time being, schoolyards are paved with asphalt concrete, with low albedo and permeability, and thus contribute to the UHI effect. The transformation of these schools into cooling islands open to the public outside of school time would therefore make up for the lack of green space in the capital (5.8 m[2]/capita[3]) and help limit urban heat in Paris, while providing climate refuges for nearby inhabitants. Paris City Hall, via its Oasis Schoolyard program, hopes to renovate all its schoolyards within the next 20 years into cool islands.

Among the first schools to undergo this kind of renovation, 10 schoolyards scheduled for works in 2020 and will be subject to scientific evaluation under the European ERDF UIA "OASIS" project, aiming among other things to quantify the microclimatic impact of the transformations. The targeted schoolyards will be equipped with meteorological station that will monitor microclimatic parameters such as temperature, heat flux, relative humidity and wind speed. The meteorological monitoring of the schoolyards before and after transformation will allow a concrete and experimental evaluation of the microclimatic impact of this initiative.

This paper will present the GIS tool used to prioritize the schoolyards under municipal management according to their cooling potential, i.e. their ability to benefit from UHI countermeasures. The indicator builds on previous work by M. Hendel et al. (Hendel et al., 2018) for urban watering in extreme heat. The indicator for cooling potential, based on morphological data, vegetative cover, and computed solar irradiance is adapted to the case of long-term, permanent UHI countermeasures.

---

[2] In 2010, out of the 553 hectares of green spaces within the city of Paris, 39 hectares were with restricted access
[3] This ratio does not take into account the Bois de Vincennes and Bois de Boulogne

## Methodology

### Identifying schoolyards using GIS

Beforehand, it is necessary to identify and map the 770+ Parisian schoolyards. In order to do so, we extracted schoolyards from the geo-referenced datasets provided by Paris City Hall: built areas, land registry[4], vegetation.

Subtracting the Built areas (listed buildings in the city) from the cadastral parcel dataset (listing the city-owned land) provides a first approximation of the schoolyards (Figure 1). A visual analysis of aerial photographs coupled with a comparison with the attributes of the other data sets as well as the layout plans provided by Paris City Hall enables us to isolate schoolyards from decorative gardens, car parks and sports grounds.

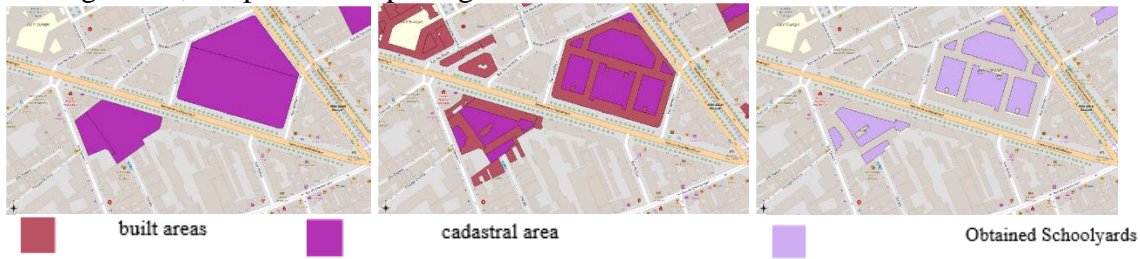

built areas      cadastral area      Obtained Schoolyards

Figure 1. From left to right: after identifying the school in the land register, then the built areas, we isolate the schoolyard by subtracting the second from the first.

### Cooling Potential Indicator

In order to estimate the cooling potential indicator, we cross-reference solar irradiance data with the schoolyards that we obtained, excluding vegetation cover.

Using a vegetation canopy model, we can identify trees and low-lying vegetation and single out bare, unshaded schoolyard areas.

### Solar Irradiance

Direct sunlight is the predominant energy source; we determine the cooling potential by evaluating the cumulative solar irradiation level of the surface of the courtyards that is not sheltered.

Solar Irradiance is calculated to be representative of a Parisian summer in clear, cloudless sky conditions. It is simulated from a digital elevation model (DEM) with a resolution of 0.5 m obtained from LIDAR measurements made in 2012.

The daily solar irradiance of an unmasked flat surface is calculated with a 30-minute time step during the heatwave-prone period that stretches from June 15 to September 15 (Figure 2).

---

[4] The cadastral parcel is the unitary element of land ownership. It is formed by any portion of land in a single block belonging to a single owner (or to the same joint ownership). The DGFiP manages the cadastral parcel. The latter is a territorial division, and is an entity contained in the same commune, and the same section. Not all the space is fragmented by the land registry (public road, river...).

The cadastral plan that we used was updated in July 2015

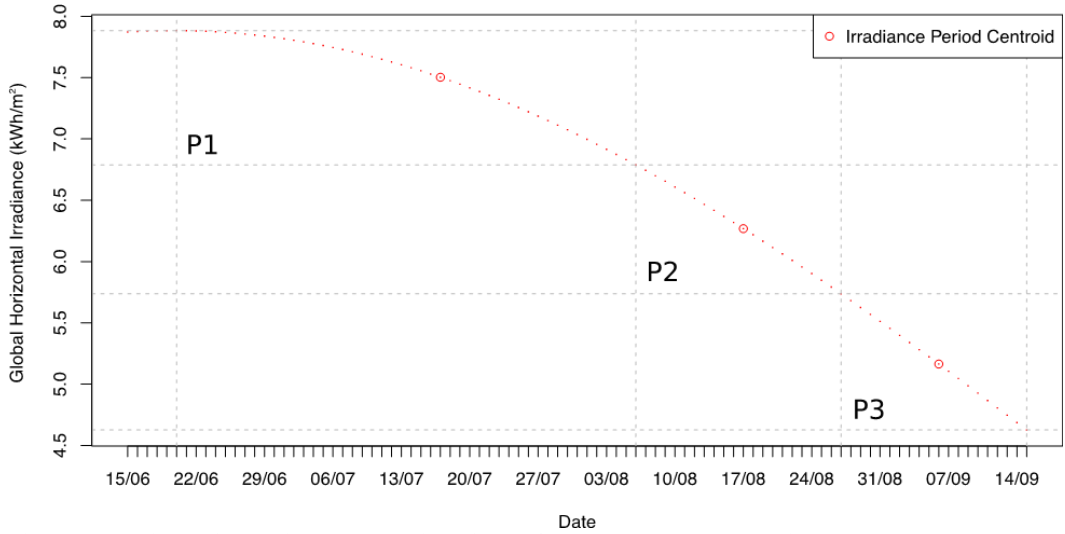

Figure 2. Global daily horizontal solar irradiance from June 15 to September 15

In order to optimize calculation time, the surface irradiance, also computed with a 30-minute time step, we select three representative days by dividing the daily overall horizontal sunshine into three energy bands: P1, P2 and P3 as defined in Table 1. The centroids of the bands correspond to the three representative dates. We thus restrict the calculations to these 3 days.

Table 1. Solar Irradiance energy intervals and representative dates

| Solar irradiance (kWh/m²) | Min | Max | Date |
|---|---|---|---|
| P1 | 7.9 | 6.8 | July 17[th] |
| P2 | 6.8 | 5.7 | August 17[th] |
| P3 | 4.6 | 5.7 | September 6[th] |

The cooling potential analysis is based on a precise time range, a target period. In the case of the schoolyards, this period extends from 6 am to 6 pm UTC (solar time), i.e. half a day. In addition, we set a critical solar irradiance threshold, above which we consider the cooling potential to be high and low below. The threshold value corresponds to half of the total energy that the surface received over the target period. Within the framework of the OASIS project, the transformations are sustainable over time and must therefore be adapted to the entire summer period. We therefore calculate the average sunshine representative of summer, derived from the three standard days.

A weighted average of the sunshine thresholds corresponding to the 3 typical days, makes it possible to obtain an overall critical threshold above which the sunny surface is declared to have a high cooling potential (see Table 2)

Table 2. Solar irradiance threshold computed for each standard day, and global irradiance threshold

| Energy band | Number of days | Solar irradiance Threshold (kWh/m²) |
|---|---|---|
| P1 | 52 | 3.6791 |
| P2 | 21 | 3.1206 |
| P3 | 19 | 2.5996 |
| Total Number of days | 92 | |
| Global Threshold | | 3.3287 |

**Existing paving materials**

Schoolyards are paved with asphalt concrete. However, they also have vegetation covers, i.e. trees, low-rise vegetation that we must take into account while evaluating the cooling potential. In order to do so, we use the raster image provided by the APUR, which is obtained by processing aerial imaging of land cover: each pixel value represents the height of the vegetation cover. We proceeded to vectorising the raster image in order to ease data handling.

**The Schoolyards' cooling potential**

Once we have identified the schoolyards, it is possible to calculate their cooling potential. At the scale of the School building, we calculate the ratio of high cooling potential areas over the total schoolyard area. This allows us to rank schoolyards accordingly and categorizing into priority levels.

Figure 3 shows the process that is undergone in order to compute the schoolyard's cooling potential. The left image shows the mean solar irradiance raster image of a schoolyard. Using the vegetation vector layer, we create a clipping mask in order to remove vegetalized areas from the mean solar irradiance raster. The center image shows in green, the vegetation mask. The raster obtained is the mean solar irradiance raster image of the bare, unshaded surface of the schoolyard. We then apply our critical threshold to the newly obtained raster, thus spotting the high cooling potential areas, i.e. the areas that would benefit the most from cooling strategies (right image).

We compute the areas of the high cooling potential surface previously obtained, and then calculate the ratio of these areas over the total area of the schoolyard. We obtain a percentage that will help us determine the priority level of the schoolyard. If less than a quarter of the schoolyard's ratio is of high cooling potential, our schoolyard has a low overall cooling potential. The overall cooling potential of the schoolyard rises according to the percentage of high cooling potential areas in the schoolyard. The level is moderate when less than half of the schoolyard's surface is of high cooling potential, medium when it rises to more than the half, and high when more than three quarters of the surface have a high cooling potential.

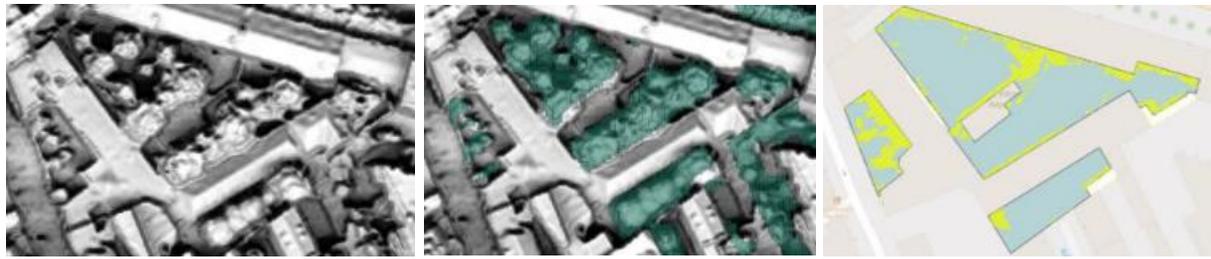

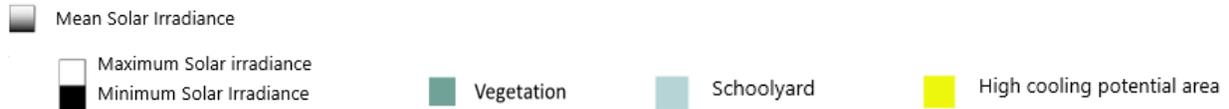

Figure 3. Method for calculating the surface ratio with high cooling potential. We apply the critical threshold to the mean solar irradiance map, while removing vegetalized areas.

## Results and discussions

Initial calculations helped create the schoolyard dataset and build a cooling potential indicator based on the proportion of schoolyards' areas with high cooling potential (Figure 4). Out of the 670 establishments managed by the City of Paris, the suggested methodology identifies 38 facilities with high cooling potential, 157 with medium potential and 286 with moderate potential.

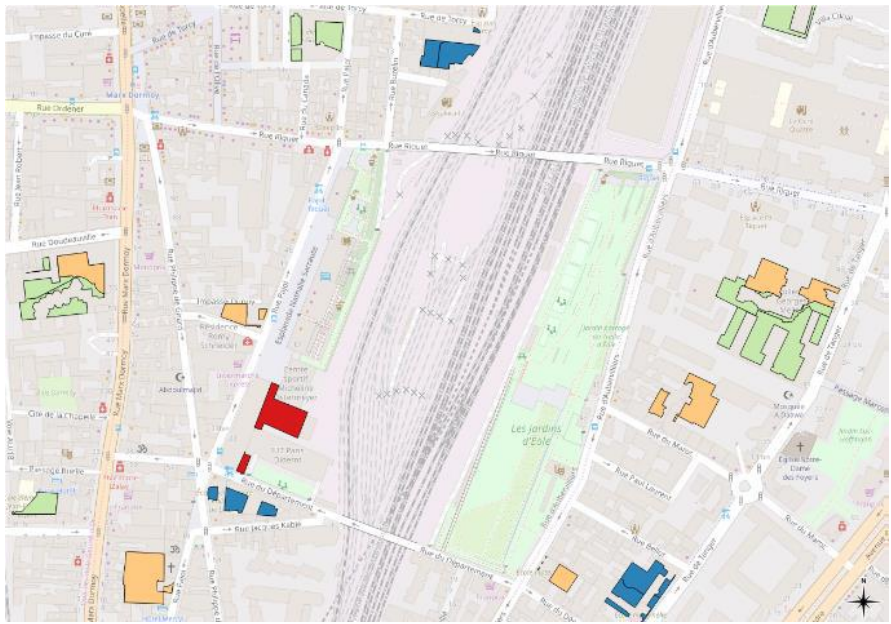

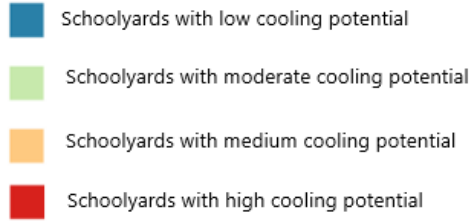

- Schoolyards with low cooling potential
- Schoolyards with moderate cooling potential
- Schoolyards with medium cooling potential
- Schoolyards with high cooling potential

Figure 4. Mapping of the cooling potential of schoolyards around the Paris Diderot IUT in Paris. The cooling potential is proportional to the percentage of the yard's surface area receiving surface energy above the previously defined sunlight threshold. Respectively, the scale represents schoolyards having at least 1%, 24%, 47% or 71% of surface area with high cooling potential.

In our case, we applied this selection criterion to schoolyards, but we could easily transpose it to other facilities (terraces, gardens, public and private spaces, etc.). The indicator is indeed intrinsic to the geometric and physical characteristics of the apprehended area. It takes into account vegetation and sunshine.

The method can be refined insofar as the cooling potential indicator currently concerns an average over the entire schoolyard attached to an establishment, without taking into account the connectedness of the spaces. We can indeed take as an example a multiple courtyard, formed by two lots, one of which is very shaded, and the other very exposed (Figure 5).

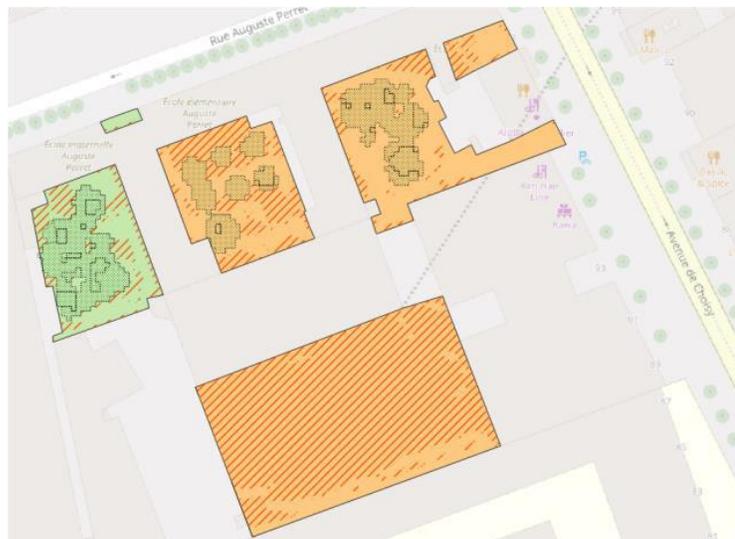

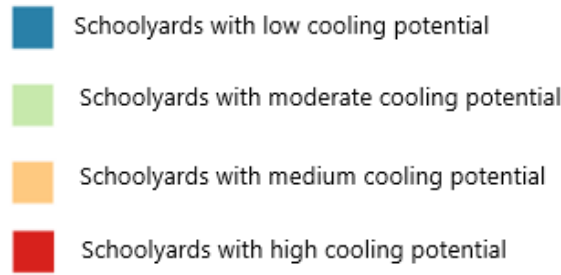

Figure 5. Example of a multiple-entity schoolyard in the Choisy neighborhood. Red lines represent high potential, while vegetation is represented as a green cloud.

## Conclusion

This preliminary work aimed to identify schoolyards with a high cooling potential by setting up an indicator. This enables us to target schoolyards where improvement of thermal stress is expect to improve drastically following the renovation work: Out of 670 facilities, we identified 38 with a high cooling potential.

Other future analyses will focus on cross-referencing this information with a heat wave risk assessment, the latter being broken down into several factors. Among the heat wave risk factors, a spatial assessment of microclimatic hazard will determine areas with record temperatures during heat wave episodes. Similarly, we will define the exposure factor to estimate the number of people affected by the heat wave, for instance, children attending the facility or the population density around the facility. Finally, an assessment of the vulnerability of populations will be integrated, quantifying their ability or not to protect themselves during a heat wave.

Future analysis will consist in cross-referencing meteorological data (including air temperature simulations from the EPICEA[5] project), with the influence of nearby parks, vegetation and water points, and then to integrate socio-economic criteria (REP, INSEE socio-economic data, etc.). The schoolyards thus selected for the OASIS project will undergo microclimatic monitoring before and after the retrofitting. The targeted schoolyards will then be equipped with meteorological station that will monitor microclimatic parameters such as temperature, heat flux, relative humidity and wind speed. These stations will help us observe the evolution of the schoolyards condition prior to the transformation and after the renovation. Other data sets include meteorological information gathered by private individuals. These datasets are interesting, as they directly involve Paris inhabitants and promote citizen participation, in addition to helping us cover larger areas for our investigation, even if the data is less precise.

The construction of this indicator is part of the process of developing a GIS tool to support decision-making, and could be applied to other city facilities, or even transposed to other cities and territories. Provided that we have, for a given city, an equivalent to land registry, (such as aerial imaging of buildings and constructions) vegetation and solar irradiance data, we can apply

---

[5] Epicea: Multidisciplinary study of the impacts of climate change on the scale of the Paris urban area, Project jointly conducted by Météo-France, the Centre Scientifique et Technique du Bâtiment (CSTB) and the City of Paris

our methodology to diverse surfaces, such as courtyards, sidewalks, roads, rooftops… and compute the high cooling potential surface areas, thus enabling us to apply our indicator.

## Acknowledgement

This study is funded by the European ERDF project UIA03-344-OASIS. The authors would also like to thank the APUR for the solar irradiance datasets.